\documentclass[12pt]{article}
\usepackage{graphicx}
\usepackage{amsmath}
\input{tcilatex}

\begin{document}
\begin{center}
{\bf \large Self-duality of the asymptotic relaxation states of fluid and plasmas}\\
\vspace{1cm}
{\bf F. Spineanu and M. Vlad} \\
{\sl Association Euratom-MER Romania\\
National Institute of Laser, Plasma and Radiation Physics, MG-36 Magurele,\\
Bucharest, Romania}\\
(\emph{e-mail: florin@drfc.cad.cea.fr, madi@drfc.cad.cea.fr})
\end{center}

\begin{abstract}
The states of asymptotic relaxation of $2$-dimensional fluids and plasma
present a high degree of regularity and obey to the $\sinh $-Poisson
equation. We find that embedding the classical fluid description into a
field-theoretical framework, the same equation appears as a manifestation of
the self-duality.
\end{abstract}

The states generated by externally driving (stirring) an ideal fluid can
have very irregular form. It is however known from experiments and numerical
simulations that after suppressing the drive the system evolves to states
with high degree of order, essentially consisting of few large vortices,
with very regular geometry. These states are attained after long time
evolution and are not due to the residual dissipation. The process consists
of vortex merging, which is an essentially topological event where the weak
dissipation only allows the reconnection of the field lines but does not
produce significant energy loss from the fluid motion. Inferring from
results of numerical simulations, Montgomery \emph{et al.} \cite{Montg1}, 
\cite{Montgomery} have proved that the scalar stream function $\psi $
describing the motion in two-dimensional space obeys in the far asymtotic
regime (where the regular structures are dominant) the \emph{sinh}-Poisson
equation 
\begin{equation}
\Delta \psi +\gamma \sinh \left( \beta \psi \right) =0  \label{sinhP}
\end{equation}
where $\gamma $ and $\beta $ are \emph{positive} constants. The relations of 
$\psi $ to the velocity and vorticity are $\mathbf{v=\nabla }\psi \times 
\widehat{\mathbf{e}}_{z}$ , $\mathbf{\omega }=\mathbf{\nabla }\times \mathbf{%
v=-}\nabla ^{2}\psi \widehat{\mathbf{e}}_{z}$ where $\widehat{\mathbf{e}}%
_{z} $ is the unitary vector perpendicular to the plane. With these
variables, the Euler equations for the two dimensional ideal incompressible
fluid are 
\begin{eqnarray}
\frac{\partial \mathbf{\omega }}{\partial t}+\left( \mathbf{v\cdot \nabla }%
\right) \mathbf{\omega } &=&\mathbf{0}  \label{Euler} \\
\mathbf{\nabla \cdot v} &=&0  \notag
\end{eqnarray}

We will try to develop a field-theoretical model of the stationary
asymptotic relaxed states, \emph{i.e.} we look for a model that could
provide a derivation of Eq.(\ref{sinhP}). This will be done progressively,
examining models ellaborated for closely related problems and collecting the
relevant suggestions that could allow us to write a Lagrangian density.

In the study of the two-dimensional Euler fluids, and in particular in
explaining the origin of Eq.(\ref{sinhP}), an important model consists of a
system of $N$ discrete vorticity filaments perpendicular on plane, having
circular transversal section of radius $a$ and carrying the vorticity $%
\omega _{i}$, $i=1,N$. A very comprehensive account of this system is given
by Kraichnan and Montgomery \cite{KraichnanMontgomery} where the
correspondence between the continuous and discrete representations of
vorticity is discussed in detail. The motion in plane of the $k$-th filament
of coordinates $\mathbf{r}_{k}\equiv \left( r_{k}^{1},r_{k}^{2}\right)
\equiv \left( x_{k},y_{k}\right) $ is given by 
\begin{equation}
\frac{dr_{k}^{i}}{dt}=\varepsilon ^{ij}\frac{\partial }{\partial r_{k}^{j}}%
\sum_{n=1,n\neq k}^{N}\omega _{n}G\left( \mathbf{r}_{k}-\mathbf{r}%
_{n}\right) \;,\;i,j=1,2\;,\;k=1,N  \label{motion}
\end{equation}
where the summation is over all the other filaments' positions $\mathbf{r}%
_{n}$, $n\neq k$, and $\varepsilon ^{ij}$ is the antisymmetric tensor in two
dimensions. As shown in Ref.\cite{KraichnanMontgomery} $G\left( \mathbf{r}%
_{k}-\mathbf{r}_{n}\right) $ can be approximated for $a$ small compared to
the space extension of the fluid, $L$, $a\ll L$, as the Green function of
the Laplacian 
\begin{equation}
G\left( \mathbf{r},\mathbf{r}^{\prime }\right) \approx -\frac{1}{2\pi }\ln
\left( \frac{\left| \mathbf{r}-\mathbf{r}^{\prime }\right| }{L}\right)
\label{Green}
\end{equation}

The statistical properties of the system of discrete vortices have been
examined using the Liouville theorem and the conservation of energy and momentum . The model consists of an equal number of positive and negative
vortices with equal absolute magnitudes $\left| \omega _{i}\right| =\left|
\omega \right| $, in contact with a thermal bath of temperature $T$. The
possibility of the generation of two supervortices of opposite
signs arises when $T$ is {\emph negative} and sufficiently large. When
the \emph{most probable state} is attained for a stationary configuration
the stream function $\psi $ is shown to verify the $\sinh $-Poisson equation
(\ref{sinhP}). These statistical considerations remain the reference
explanation for the appearence of this equation in this context.

Highly ordered states have been obtained by numerical simulations for other
systems in two spatial dimensions: systems of filaments of electric current
in $2$-dimensional MHD, guiding centre particles. The asymptotic states are
found to be described by scalar functions obeying Liouville equation or $%
\sinh $-Poisson equation. The theoretical approach which permitted to
explain this result was based on the extremum of the entropy over a
statistical ensemble of states of a set of discrete objects under the
constraints which express the conservation of the invariants of motion and
with a negative temperature.

In the equations of motion (\ref{motion}) the right hand side contains the 
\emph{curl} of the Laplacian Green's function (\ref{Green}) (we take $L=1$) 
\begin{equation}
-\varepsilon ^{ij}\partial _{j}G\left( \mathbf{r,r}^{\prime }\right)
=\varepsilon ^{ij}\partial _{j}\frac{1}{2\pi }\ln r=\frac{1}{2\pi }%
\varepsilon ^{ij}\frac{r^{j}}{r^{2}}  \label{avect}
\end{equation}
\begin{equation*}
\mathbf{\nabla }^{2}\frac{1}{2\pi }\ln r=\delta ^{2}\left( r\right)
\end{equation*}
The term in right hand side of (\ref{motion}) can be seen as a vector
potential $\mathbf{a}\left( \mathbf{r},t\right) $. The ``magnetic'' field $%
\mathbf{\nabla \times a}$ is a sum of Dirac $\delta $ functions at the
locations of the vortices. If we take equal strength $\omega $ for all
vortices this potential appears as the ``statistical potential'' and has a
topological interpretation \cite{0010246}. From (\ref{avect}) it can be
rewritten 
\begin{equation}
\frac{1}{2\pi }\varepsilon ^{ij}\frac{r^{j}}{r^{2}}=-\frac{1}{2\pi }\partial
_{i}\arctan \frac{y}{x}=-\frac{1}{2\pi }\partial _{i}\theta  \label{statist}
\end{equation}
where $\mathbf{r=}\left( x,y\right) =\left( r\cos \theta ,r\sin \theta
\right) $. The ``magnetic'' flux through a surface limitted by a large
circle is proportional with the number of vortices. The topological nature
of this potential suggests that it can be naturally derived in a topological
framework, \emph{i.e.} from a Lagrangian density of the Chern-Simons type, 
\begin{equation}
\mathcal{L}=\frac{1}{4}\varepsilon ^{\mu \alpha \beta }A_{\nu }F_{\alpha
\beta }  \label{CSdens}
\end{equation}
where $F_{\alpha \beta }=\partial _{\alpha }A_{\beta }-\partial _{\beta
}A_{\alpha }$. In the analysis for the two-dimensional Euler equation (as in
the investigation of similar systems), the vorticity-type functions $\mathbf{%
\omega }$ are expressed as the Lapacian operator applied on a scalar
function and this naturally invokes the Green function of the Laplacian.
Then this approach suggests that the intrinsically determined motion of the
fluid can be projected onto two distinct parts of a new model: objects
(vortices) and interaction between them (the potential obtained from the
Green function). The fact that the interaction potential can be derived from
a Chern-Simons topological action suggests that, in order to embedd the
original Euler fluid system into a larger field theoretical context, we need
(a) a ``matter'' part in the Lagrangian, which should provide the free
dynamics of the vortices; (b) the Chern-Simons term, to describe the free
dynamics of the field; (c) the interaction term of the (Chern-Simons-) field
and the matter. The main requirement to this field-theoretical extension of
the original Euler fluid model is to \ reproduce the discrete vortices and
their equation of motion.

Jackiw and Pi \cite{JackiwPi}, \cite{JakPi90} have examined a model of $N$
interacting particles (of charges $e_{s}$) moving in plane described by the
Lagrangian 
\begin{equation}
L=\sum_{s=1}^{N}\frac{1}{2}m_{s}\mathbf{v}_{s}^{2}+\frac{1}{2}\int
d^{2}r\varepsilon ^{\alpha \beta \gamma }\left( \partial _{\alpha }A_{\beta
}\right) A_{\gamma }-\int d^{2}rA_{\mu }j^{\mu }  \label{LagJP}
\end{equation}
where $m_{s}\mathbf{v}_{s}=\mathbf{p}_{s}-e_{s}\mathbf{A}\left( \mathbf{r}%
_{s}|\mathbf{r}_{1},\mathbf{r}_{2},...,\mathbf{r}_{N}\right) $, $\mathbf{A}%
\left( \mathbf{r}_{s}|\mathbf{r}_{1},\mathbf{r}_{2},...,\mathbf{r}%
_{N}\right) \equiv (a_{s}^{i}\left( \mathbf{r}_{1},\mathbf{r}_{2},...,%
\mathbf{r}_{N}\right) _{i=1,2}$ and $a_{s}^{i}\left( \mathbf{r}_{1},\mathbf{r%
}_{2},...,\mathbf{r}_{N}\right) =\frac{1}{2\pi }\varepsilon ^{ij}\sum_{q\neq
s}^{N}e_{q}\frac{r_{s}^{j}-r_{q}^{j}}{\left| \mathbf{r}_{s}-\mathbf{r}%
_{q}\right| ^{2}}$. The matter current is defined as $j^{\mu }\equiv \left(
\rho ,\mathbf{j}\right) =\sum_{s=1}^{N}e_{s}v_{s}\delta \left( \mathbf{r-r}%
_{s}\right) $, $v_{s}^{\mu }=(1,\mathbf{v}_{s})$ with the metric $\left(
1,-1,-1\right) $. By varying the action we obtain the following equations 
\begin{equation*}
\frac{1}{2}\varepsilon ^{\alpha \beta \gamma }F_{\alpha \beta }=\varepsilon
^{\gamma \alpha \beta }\partial _{\alpha }A_{\beta }=j^{\gamma }
\end{equation*}
\begin{equation}
B=-\rho  \label{beqrho}
\end{equation}
\begin{equation}
E^{i}=\varepsilon ^{ij}j^{j}  \label{ej}
\end{equation}
Here $B=\varepsilon ^{ij}\partial _{j}A_{i}$. In these equations the
potential appears as being generated by the matter, \emph{i.e.} by the
``current'' of particles. We note that the Eq.(\ref{beqrho}) connects the
matter density with the curl of the potential. This is important since in
order to derive it in the context of the quantum version of their model,
Jackiw and Pi have shown that one needs to include a nonlinear
self-interaction of the wave function representing the matter field. The
self-interaction of the scalar matter field is of the type $\varphi ^{4}$.
The canonical momentum expression becomes the covariant derivative in the
field theory version. Even at this point where we only have some hints, it
is suggestive to look for possible identifications between the fluid
variables and the field theoretical variables of the model of Jackiw and Pi.
The fluid variables are: the scalar potential $\psi $, the velocity $%
v^{i}=\varepsilon ^{ij}\partial _{j}\psi $ and the vorticity $\omega
=\varepsilon ^{ij}\partial _{i}v_{j}$. In order to check the possibility
that the fluid model can be embedded into the field theory just described we
identify 
\begin{equation}
\omega \equiv \rho =\Psi ^{\ast }\Psi  \label{omegarho}
\end{equation}
\begin{equation}
v^{i}\equiv A^{i}  \label{va}
\end{equation}
Consider the relation between the vector potential $\mathbf{A}$ and the
density $\rho $, obtained from Eq.(\ref{beqrho}) 
\begin{equation*}
\partial ^{i}\partial _{i}A^{j}=-\varepsilon ^{jk}\partial _{k}\rho
\end{equation*}
or, writting symbolically the Green function of the Laplace operator for
argument $\mathbf{r-r}^{\prime }$ as the inverse of the operator at the
left, we have 
\begin{equation}
A^{j}\left( \mathbf{r},t\right) =\varepsilon ^{jk}\partial _{k}\int
d^{2}r^{\prime }\left( -\partial ^{i}\partial _{i}\right) _{rr^{\prime
}}^{-1}\rho \left( \mathbf{r}^{\prime },t\right)  \label{Arho}
\end{equation}
Since $\omega =-\nabla ^{2}\psi $ (where $\nabla ^{2}\equiv \partial
^{i}\partial _{i}$) we have formally $\psi =-\left( \nabla ^{2}\right) ^{-1}$
$\omega $ and use this formula to calculate the right side term of (\ref
{Arho}) taking into account the identification (\ref{omegarho}) 
\begin{equation*}
\varepsilon ^{jk}\partial _{k}\int d^{2}r^{\prime }\left( -\partial
^{i}\partial _{i}\right) _{rr^{\prime }}^{-1}\omega \left( \mathbf{r}%
^{\prime }\right) =\varepsilon ^{jk}\partial _{k}\psi =v^{j}
\end{equation*}
which confirms (\ref{va}). It results that if we assume the identification (%
\ref{omegarho}) the equation (\ref{beqrho}) obtained form the Lagrangian is
precisely the \emph{definition }of the vorticity vector.

One important hint from the model of Jackiw and Pi is the idea to represent
a classical quantity as the modulus of a fictitious complex scalar field, as
in Eq.(\ref{omegarho}) and derive dynamical equations from a Lagrangian
density expressed in terms of this field. The substitution of the dynamics
expressed in terms of usual mechanical quantities by the richer dynamics of
the amplitude and phase of the complex scalar field is an example of
embedding of one theory into a larger framework and relies on the example of
quantum mechanics.

In Ref.(\cite{Nardelli}) the $\sinh $-Poisson equation is derived in a field
theoretical model where instead of the topological coupling two complex
scalar fields are considered. In that model the dynamics of the two scalar
fields is independent except that the self-interaction depends on both.
Combining this suggestion with the one exposed in the previous paragraph, we
will take the density $\rho $ (which we identify with the local value of the
vorticity $\omega $ as in Eq.(\ref{omegarho}) ) as a two-component complex
field $\Psi $ 
\begin{equation*}
\rho \sim \left[ \Psi ^{\dagger },\Psi \right]
\end{equation*}
where $\Psi =\left( 
\begin{array}{c}
\phi _{1} \\ 
\phi _{2}
\end{array}
\right) $. Keeping the same structure of the Lagrangian density as in the
previous example but extending to a non-Abelian $SU\left( 2\right) $ gauge
field we have \cite{Dunne}, \cite{JakPi90} 
\begin{equation}
\mathcal{L}=-\varepsilon ^{\mu \nu \rho }tr\left( \partial _{\mu }A_{\nu
}A_{\rho }+\frac{2}{3}A_{\mu }A_{\nu }A_{\rho }\right) +itr\left( \Psi
^{\dagger }D_{0}\Psi \right) -\frac{1}{2}tr\left( \left( D_{i}\Psi \right)
^{\dagger }D_{i}\Psi \right) +\frac{1}{4}tr\left( \left[ \Psi ^{\dagger
},\Psi \right] \right) ^{2}  \label{LagSU2}
\end{equation}
where the potantial $A$ takes values in the algebra of the group $SU\left(
2\right) $. This form incorporates and adapt all the suggestions from the
previous models: (a) the first term is the general non-Abelian expression of
the Chern-Simons term; (b) it uses the covariant derivatives ($\mu =0,1,2$, $%
i=1,2$) for the minimal coupling 
\begin{equation*}
D_{\mu }\Psi =\partial _{\mu }\Psi +\left[ A_{\mu },\Psi \right]
\end{equation*}
(c) Finally, it includes a scalar self-interaction of $\varphi ^{4}$ type.
Analogous to the case treated by Jackiw and Pi for the $U\left( 1\right) $
gauge field (\ref{LagJP}), in Ref.\cite{Dunne} it is shown that the
Hamiltonian density corresponding to the Lagrangian density (\ref{LagSU2})
is 
\begin{equation}
\mathcal{H}=\frac{1}{2}tr\left( \left( D_{i}\Psi \right) ^{\dagger }\left(
D_{i}\Psi \right) \right) -\frac{1}{4}tr\left( \left[ \Psi ^{\dagger },\Psi 
\right] ^{2}\right)  \label{HamiltSU2}
\end{equation}
since the Chern-Simons term does not contribute to the energy density (being
first order in the time derivatives). The equations of motion are 
\begin{equation}
iD_{0}\Psi =-\frac{1}{2}\mathbf{D}^{2}\Psi -\frac{1}{2}\left[ \left[ \Psi
,\Psi ^{\dagger }\right] ,\Psi \right]  \label{ec1}
\end{equation}
\begin{equation}
F_{\mu \nu }=-\frac{i}{2}\varepsilon _{\mu \nu \rho }J^{\rho }  \label{ec2}
\end{equation}
Using the notation $D_{\pm }\equiv D_{1}\pm iD_{2}$ the first term in (\ref
{HamiltSU2}) can be written 
\begin{equation*}
tr\left( \left( D_{i}\Psi \right) ^{\dagger }\left( D_{i}\Psi \right)
\right) =tr\left( \left( D_{-}\Psi \right) ^{\dagger }\left( D_{-}\Psi
\right) \right) +\frac{1}{2}tr\left( \Psi ^{\dagger }\left[ \left[ \Psi
,\Psi ^{\dagger }\right] ,\Psi \right] \right)
\end{equation*}
The last term comes from Eq.(\ref{ec2}) and from the definition $J^{0}=\left[
\Psi ,\Psi ^{\dagger }\right] $. Then the energy density is 
\begin{equation}
\mathcal{H}=\frac{1}{2}tr\left( \left( D_{-}\Psi \right) ^{\dagger }\left(
D_{-}\Psi \right) \right) \geq 0  \label{Bogo}
\end{equation}
and the Bogomol'nyi inequality is saturated at \emph{self-duality} 
\begin{equation}
D_{-}\Psi =0  \label{sd1}
\end{equation}
\begin{equation}
\partial _{+}A_{-}-\partial _{-}A_{+}+\left[ A_{+},A_{-}\right] =\left[ \Psi
,\Psi ^{\dagger }\right]  \label{sd2}
\end{equation}
The first equation results from the minimum in Eq.(\ref{Bogo}) and the
second is actually Eq.(\ref{ec2}) with the definitions 
\begin{eqnarray*}
J^{0} &=&\left[ \Psi ^{\dagger },\Psi \right] \\
J^{i} &=&-\frac{i}{2}\left( \left[ \Psi ^{\dagger },D_{i}\Psi \right] -\left[
\left( D_{i}\Psi \right) ^{\dagger },\Psi \right] \right)
\end{eqnarray*}
The \emph{static} solutions of the \emph{self-duality} equations (\ref{sd1}, 
\ref{sd2}) are derived in \cite{Dunne}, using the algebraic \emph{ansatz}: 
\begin{equation*}
A_{i}=\sum_{a=1}^{r}A_{i}^{a}H_{a}
\end{equation*}
\begin{equation*}
\Psi =\sum_{a=1}^{r}\psi ^{a}E_{a}+\psi ^{M}E_{-M}
\end{equation*}
where $H_{a}$ are the Cartan subalgebra generators for the gauge Lie
algebra, $E_{a}$ are the simple-root step operators and $E_{-M}$ is the step
operator corresponding to minus the maximal root \cite{Sattinger}. The rank
of the algebra is noted $r$, and $r=1$ for $SU\left( 2\right) $. Then 
\begin{equation}
\left[ \Psi ^{\dagger },\Psi \right] =\sum_{a=1}^{r}\left| \psi ^{a}\right|
^{2}H_{a}+\left| \psi ^{M}\right| ^{2}H_{-M}  \label{psidagpsi}
\end{equation}
The equations (\ref{sd1}, \ref{sd2}) lead to the affine Toda equations 
\begin{equation}
\nabla ^{2}\ln \rho _{a}+\sum_{b=1}^{r+1}\widetilde{C}_{ab}\rho _{b}=0
\label{Toda}
\end{equation}
for $a=1,r$, plus the index for $M$, \emph{i.e.} $a=1,2$. $\widetilde{C}%
_{ab} $ is the extended Cartan matrix 
\begin{equation*}
\widetilde{C}_{ab}=\frac{2\mathbf{\alpha }^{\left( a\right) }\cdot \mathbf{%
\alpha }^{\left( b\right) }}{\left| \mathbf{\alpha }^{\left( b\right)
}\right| ^{2}},\;a,b=1,2
\end{equation*}
where $\mathbf{\alpha }^{\left( a\right) }$ are the simple root vectors of
the algebra $su\left( 2\right) $, and in addition the minus maximal root.
The equations (\ref{Toda}) can be written in detail for $\rho _{1}\equiv
\left| \psi ^{1}\right| ^{2}$, $\rho _{2}\equiv \left| \psi ^{-M}\right|
^{2} $ 
\begin{eqnarray}
\Delta \ln \rho _{1}+2(\rho _{1}-\rho _{2}) &=&0  \label{deltarho12} \\
\Delta \ln \rho _{2}+2(-\rho _{1}+\rho _{2}) &=&0  \notag
\end{eqnarray}
and this gives the relation 
\begin{equation*}
\Delta \ln \left( \rho _{1}\rho _{2}\right) =0
\end{equation*}
or 
\begin{equation}
\rho _{2}=\text{const}\;\rho _{1}^{-1}  \label{rho12}
\end{equation}
since the exponential of a linear term is excluded by the conditions on a
circle at infinity. Using Eq.(\ref{psidagpsi}) we have to identify 
\begin{equation}
\omega =\rho _{1}-\rho _{2}  \label{omrho1min2}
\end{equation}
whose equation is obtained from equations in (\ref{deltarho12}) using (\ref
{rho12}) const$=1$ (see bellow) 
\begin{equation}
\Delta \ln \rho _{1}+2(\rho _{1}-\rho _{1}^{-1})=0  \label{ecrho1}
\end{equation}
The substitution $\left| \beta \right| \psi \equiv \ln \rho _{1}$ leads both
equations (\ref{omrho1min2}) and (\ref{ecrho1}) to the form 
\begin{equation}
\Delta \psi +\frac{4}{\left| \beta \right| }\sinh \left( \left| \beta
\right| \psi \right) =0  \label{happy}
\end{equation}

Choosing a constant different of unity in Eq.(\ref{rho12}) imposes a linear
substitution of the stream function : $\psi \rightarrow \psi ^{\prime
}\equiv \gamma +\beta \psi $ and modifies the factor multiplying the $\sinh $
function in Eq.(\ref{happy}).

We have constructed the Lagrangian density (\ref{LagSU2}) with a standard structure of matter-gauge field interaction and incorporating suggestions from models relevant for our objective. The aim was to reproduce the model of fluid vortex filaments supposed to be a faithfull representation of the original Euler fluid dynamics.
The fundamental step in the derivation of the $\sinh $-Poisson equation is
the assumption that the energy density (\ref{Bogo}) is minimum (which is
equivalent to minimizing the \emph{action} for these stationary solution) 
\emph{i.e.} in the condition of \emph{self-duality}.

We conclude that the Euler fluid equations can be embedded into a field
theoretical framework via the discrete model of vortex filaments. This
framework is able to provide the equation obeyed by the scalar stream
function at stationarity and this is precisely the equation inferred from
numerical experiments. The particularity of the field theoretical framework
is that it shows that the asymptotic states consisting of regular vortices
correspond to \emph{self-dual} states in field theory. This might have deep
significance and deserves further investigation.

One must also note that the statistical-theoretical model able to explain
the $\sinh $-Poisson equation \cite{Montg2}, \cite{Montg3}, \cite{Joyce}, 
\cite{Smith} appears to have strong and deep connections
with the self-duality. It should be carefully considered as a possible
approach in the study of models where self-duality is known to exist, like $%
O\left( n\right) $ or Self-Dual Yang-Mills theory.

\textbf{Aknowledgement}. This work has been presented at the \emph{Forum of
Theory} organized by the Department de Recherche sur la Fusion Controlee,
CEA-Cadarache, in Aix-en-Provence, France (July 2001). The authors are
grateful to the organizers for inviting them to this meeting.

\end{document}